\documentclass[conference,a4paper]{IEEEtran}
\IEEEoverridecommandlockouts
\usepackage{cite}
\usepackage{amsmath,amssymb,amsfonts}
\usepackage{graphicx}
\usepackage{textcomp}
\usepackage{xcolor}

\usepackage[linesnumbered, ruled]{algorithm2e}
\SetAlFnt{\small}
\SetAlCapFnt{\small}
\SetAlCapNameFnt{\small}
\SetAlgoNlRelativeSize{-1}

\SetKwRepeat{Do}{do}{while}%

\usepackage{setspace}

\renewcommand{\IEEEbibitemsep}{0pt plus 0.5pt}
\makeatletter
\IEEEtriggercmd{\reset@font\normalfont\fontsize{7pt}{7pt}\selectfont}
\makeatother
\IEEEtriggeratref{1}

\newcount\Comments  
\usepackage{color}
\definecolor{blue}{rgb}{0,0,1}
\definecolor{red}{rgb}{1,0,0}
\newcommand{\kibitz}[2]{\ifnum\Comments=1\textcolor{#1}{#2}\fi}
\newcommand{\ke}[1]  {\kibitz{blue}   {[KE: #1]}}

\newcommand{\e}[1]  {  {\bigl\{#1\bigr\}_{\mathcal{E}}}}

\usepackage{mathtools}  

\usepackage{tabulary}
\usepackage{booktabs}

\usepackage{multirow}

\usepackage[
    n, lambda,
    operators,
    advantage,
    sets,
    adversary,
    landau,
    probability,
    notions,    
    logic,
    ff,
    mm,
    primitives,
    events,
    complexity,
    asymptotics,
    keys]{cryptocode}

    \createpseudocodeblock{pcb}{center,boxed}{}{}{}
\usepackage{comment}

\usepackage{booktabs} 

\usepackage{graphicx}
\usepackage{slashbox}

\usepackage{comment}
\usepackage{caption}

\usepackage{subcaption,graphicx}

\usepackage{tikz}

\def\BibTeX{{\rm B\kern-.05em{\sc i\kern-.025em b}\kern-.08em
    T\kern-.1667em\lower.7ex\hbox{E}\kern-.125emX}}

\begin{document}
	
	\title{A Privacy-Preserving and Accountable Billing Protocol for Peer-to-Peer Energy Trading Markets  \\
	
 \thanks{This work was supported by EPSRC through EnnCore [EP/T026995/1] and by the Flemish Government
through FWO-SBO SNIPPET project [S007619]. K.E is funded by The Ministry of National Education, Republic of Turkey.}
	}

	\author{
		\IEEEauthorblockN{
			Kamil Erdayandi\IEEEauthorrefmark{1}, Lucas C. Cordeiro\IEEEauthorrefmark{1} and 
			Mustafa A. Mustafa\IEEEauthorrefmark{1}\IEEEauthorrefmark{2}}
		\IEEEauthorblockA{
			\IEEEauthorrefmark{1}\textit{Department of Computer Science}, \textit{The University of Manchester}, UK\\
			\IEEEauthorrefmark{2}\textit{imec-COSIC}, \textit{KU Leuven}, Belgium \\
			Email:  \{kamil.erdayandi, lucas.cordeiro, mustafa.mustafa\}@manchester.ac.uk 
		}
		
	}

	\maketitle

	\begin{abstract}
  
	   This paper proposes a privacy-preserving and accountable billing (PA-Bill) protocol for trading in peer-to-peer energy markets, addressing situations where there may be discrepancies between the volume of energy committed and delivered. 
    Such discrepancies can lead to challenges in providing both privacy and accountability while maintaining accurate billing. To overcome these challenges, a universal cost splitting mechanism is proposed that prioritises privacy and accountability. It leverages a homomorphic encryption cryptosystem to provide privacy and employs blockchain technology to establish accountability. 
    A dispute resolution mechanism is also introduced to minimise the occurrence of erroneous bill calculations while ensuring accountability and non-repudiation throughout the billing process. Our evaluation demonstrates that PA-Bill offers an effective billing mechanism that maintains privacy and accountability in peer-to-peer energy markets utilising a semi-decentralised approach.

        \end{abstract}
	
	\begin{IEEEkeywords}
		Billing, Privacy, Accountability, Peer-to-peer Energy Market, Homomorphic Encryption, Blockchain
	\end{IEEEkeywords}

\section*{Nomenclature}

\addcontentsline{toc}{section}{Nomenclature}
\begingroup
\footnotesize
\begin{spacing}{1.20}
\begin{IEEEdescription}[\IEEEusemathlabelsep\IEEEsetlabelwidth{$\pi_{P2P},\pi_{FiT}, \pi_{RT}$}]

\item[$c_i$, $p_j$, $u_k$]  $i$-th consumer , $j$-th prosumer, $k$-th user 

\item[$N_C$ , $N_P$, $N_U$]  Number of consumers, prosumers, users  \
 
\item[$V^{P2P}$ ] P2P market's traded electricity volume  array
\item[$V^{Real}$ ] Real electricity consumption array  
\item[$\pi_{P2P}, \pi_{FiT}, \pi_{RT} $] P2P, FiT, Retail price 
\item[$Stat$] Array of the statements of the users

\item[$Bal_{sup}$]  Balances of the supplier
\item[$inDev$] Array of the individual deviations of the users

\item[$Dev^{Tot}$] Total deviations of the users

\item[$KGen_{pe}(k)$]  Paillier key generation method 
\item[$PK_{sup}$ ,    $SK_{sup}$] Public, Private (Secret) key pair of Supplier  
    
\item[$\{.\}_{\mathcal{E}}$] Data homomorphically encrypted with $PK_{sup}$. 
\item[ $H(.)$] Hash Function 


\end{IEEEdescription}
\end{spacing}
\endgroup

\section{Introduction}

\subsection{Motivation and Background}
	Peer-to-peer (P2P) energy trading 
 enables users to obtain clean energy at more reasonable prices than traditional suppliers, making it accessible to a wider society~\cite{soto2021peer}. It facilitates direct energy exchange between households that harness renewable energy sources (RES)~\cite{tushar2020peer}. This approach empowers individuals to become active participants in the energy system~\cite{grvzanic2022prosumers}, allowing RES owners to optimise their profits and reduce their bills through trading with other users~\cite{erdayandi2022privacy}.


	Although P2P energy trading markets offer various benefits, some challenges hinder their widespread adoption. Firstly, the vast amount of data exchanged can reveal sensitive information about users~\cite{ding2020secure}, such as their energy usage habits and lifestyle patterns. Access to this data poses significant privacy risks~\cite{ramokapane2022privacy} and could potentially violate privacy protection regulations, e.g., GDPR~\cite{REGULATI80:online}. Thus, it is crucial to ensure privacy-preserving data processing and protect data from unauthorised access~\cite{montakhabi2020sharing}. Secondly, such markets require secure and accountable solutions. 
 However, it is challenging to audit transactions without a tamper-proof system~\cite{moniruzzaman2023blockchain}. To ensure fair and accurate energy trading, it is also essential to guarantee integrity and verifiability of any data used. Thirdly, often what users commit at P2P markets deviates from what they deliver due to intermittent RES output. Hence, any billing models will need mechanisms to deal with such deviations.  

\subsection{Relevant Literature}

Within P2P energy trading, two crucial phases are market clearance and billing \& settlement~\cite{abidin2020poster}. Since privacy-preserving market clearing mechanisms have already been explored~\cite{funk2022privacy, erdayandi2022privacy, abidin2016mpc}, this paper focuses on the billing phase. 

Madhusudan et al.~\cite{madhusudan2022billing} propose four billing models for P2P energy markets which account for deviations in energy volumes from the users' bids and incorporate individual, social, or universal cost-sharing mechanisms to ensure cost-effectiveness for both consumers and prosumers. Nonetheless, they do not explore user privacy. 
A privacy-preserving billing protocol that incorporates an individual cost-sharing mechanism has been proposed in~\cite{thandi2022privacy}. 
However, it relies on a remote server for bill calculations, which poses a risk of a single point of failure. 

Singh et al.~\cite{singh2021blockchain} propose a method that uses blockchain and homomorphic schemes to protect the confidentiality of user data while enabling efficient data analysis. They do not explore any billing mechanisms. Gür et al.~\cite{gur2019blockchain} propose a system based on blockchain technology and IoT devices to facilitate billing. To ensure data confidentiality, the system employs session keys and stores the encrypted data on the blockchain. However, this is still vulnerable to breaches as unauthorised parties can gain access to these keys, enabling them to access sensitive data. 

In summary, no prior study on P2P market billing fully satisfies the three essential requirements: protecting user privacy, maintaining strong system accountability, and accommodating variations in user consumption. Neglecting any of these elements undermines the market trust, transparency and fairness, which are essential to their success and sustainability. Furthermore, integrating these three features within a single platform efficiently poses considerable challenges.

\subsection{Contributions and Organization}


 To address the issues raised in the existing literature, we propose a novel privacy-preserving and accountable billing (PA-Bill) protocol, which effectively mitigate the challenges surrounding security, privacy, accountability, and user consumption variations prevalent in current studies. 
 PA-Bill utilises a universal cost-splitting billing model that mitigates the risk of sensitive information leakage due to individual deviations. It also avoids a single point of failure by performing most calculations locally in a semi-decentralised manner. To preserve privacy, the mechanism employs homomorphic encryption in bill calculations. Moreover, PA-Bill utilises blockchain technology to integrate accountability mechanisms that addresses possible conflicts during the billing calculation process. To minimise privacy leakage, only the hashed version of the data is stored on the blockchain. Finally, PA-Bill can support large communities of $500$ households. 
 
 Unlike other solutions, PA-Bill integrates privacy protection, accountability, and accommodating user consumption variations into a single solution in an efficient way. To the best of our knowledge, no previous work has successfully implemented an efficient billing model that simultaneously preserves privacy, ensures accountability, and effectively handles discrepancies between committed and delivered volume. 

 The rest of the paper is structured as follows: Section \ref{section:preliminaries} outlines the preliminaries. The proposed PA-Bill is presented in Section \ref{section:PA-Bill}. The security analysis of PA-Bill is presented in Section \ref{section:security_privacy}, while its performance is evaluated in Section \ref{section:performance_evaluation}. Finally, Section~\ref{conlcusion_future_work} concludes the paper.

\section{Preliminaries}\label{section:preliminaries}
	
	\subsection{System Model }

	\begin{figure}[!t]
		\centering
		\includegraphics[width=0.30\textwidth]{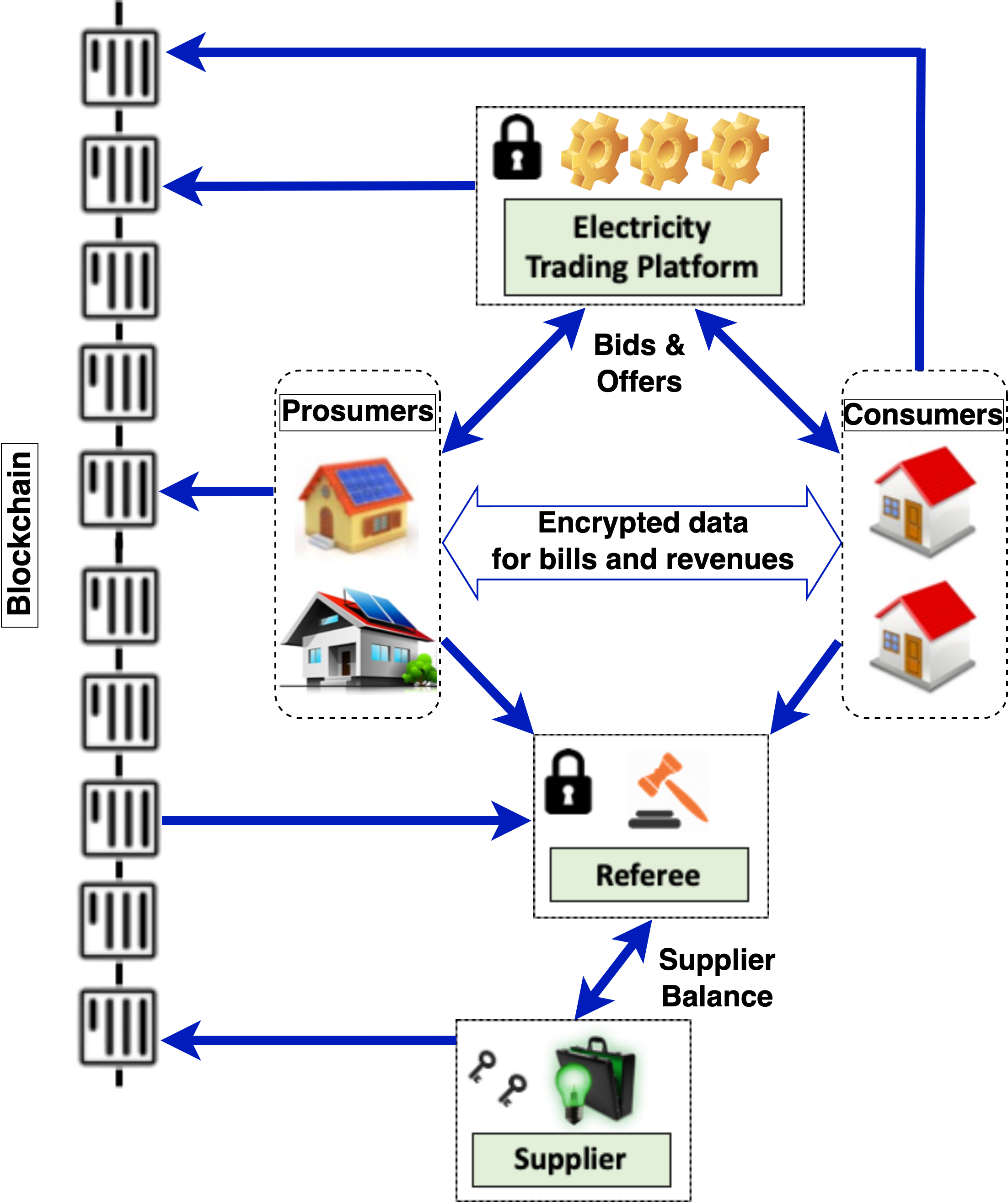}
		\caption{System model.  }
		\label{fig: System Model}
	\end{figure}

Our proposed billing protocol, illustrated in Fig.~\ref{fig: System Model}, involves prosumers, consumers, a trading platform (TP), a distributed ledger/Blockchain (DLT), a referee, and a supplier.
Prosumers generate energy through renewables, consume the volume they require, and sell any surplus energy. Consumers solely consume energy. Households have home energy management systems (HEMs) and smart meters (SMs) that measure electricity flows, provide real-time measurements, and facilitate P2P trading for the user. 
Prosumers and consumers can trade electricity through a P2P market using a trading platform~(TP). If necessary, they can also buy or sell electricity from/to a supplier as a backup option. However, P2P trading is more beneficial than relying on the supplier due to pricing considerations~\cite{erdayandi2022privacy}. Financial reconciliation occurs during settlement cycles (SCs) for users involved in trading. Within each SC, data regarding the actual electricity usage of households and their commitments to trade in the market are stored on DLT. Households calculate their bills locally in a decentralised manner. If a dispute arises, a referee intervenes to resolve it by requesting data from households and retrieving it from DLT.

	\subsection{Threat Model and Assumptions } 


       Our threat model comprises untrustworthy and semi-honest entities. Prosumers and consumers who may attempt to violate the protocol specifications and obtain sensitive data of other users are considered to be untrustworthy. Prosumers may try to maximise their revenue, while consumers may aim to minimise their expenses. Semi-honest entities include the TP, referee, and supplier. They adhere to the protocol specifications, but they may still be curious to learn sensitive data of users.

        SMs are tamper-proof and sealed. Anyone, including their users, can not tamper with them without being detected. 
        Users act rationally by seeking the most cost-effective electricity to buy or sell~\cite{abidin2016towards}. 
        We assume that the entities communicate over secure and authentic communication channels.

        \subsection{Design Requirements} 
        \label{subsection:requirement}

	\begin{itemize}
	    \item No single point of failure (SPF): 
            To avoid SPF, calculations and data storage should be distributed~\cite{erdayandi2022towards}. 
          
            \item Privacy: 
            Confidentiality of individual users' volumes of energy traded and consumed as well as individual deviation and deviation sign should be provided. 
           
            \item Accountability: Disputes arising from erroneous bill calculations must be addressed in an accountable way to prevent any party from denying responsibility.

            \item Fair deviation cost distribution: cost of P2P market deviation should be split fairly among market participants.  
            
	\end{itemize}


	\subsection{Building Blocks}

Homomorphic encryption (HE) enables computations to be performed on encrypted data, resulting in encrypted outputs that produce the same results as if the operations were conducted on unencrypted data~\cite{yi2014homomorphic}. Specifically, we deploy the Paillier cryptosystem which supports homomorphic addition and scalar multiplication on ciphertexts~\cite{paillier1999public}. Our solution ensures the privacy of households by encrypting sensitive information such as energy consumption data per SC. Billing calculations are performed on this encrypted data, thereby preserving the confidentiality of the information.
We use blockchain technology to provide accountability by ensuring that transactions are permanently recorded in a decentralised and immutable system with append-only storage. Transactions recorded on a blockchain cannot be altered by design, ensuring that they are accurate and trustworthy~\cite{singh2021blockchain}.

	\section{Privacy Preserving and Accountable Billing (PA-Bill) Protocol} 
	\label{section:PA-Bill}


In this section, we propose a privacy-preserving and accountable billing protocol for P2P energy market where users' actual energy consumption may differ from the volumes they committed. It protects sensitive household information and enables system entities to verify accurate billing calculations. 

\subsection{PA-Bill Overview}
\label{subsection:PA_overview}

The process of PA-Bill protocol is illustrated in Fig.~\ref{fig: System Model}, which includes interactions between the entities. The system utilises the public-private key pair of the supplier for all homomorphically encrypted calculations. A distinct set of HE keys, namely $PK_{sup}$ and $SK_{sup}$ are generated for each billing month. Additionally, each month the consumers and prosumers are paired together to perform accountable calculations.

In the energy trading model, users send homomorphically encrypted bid-offer data to the TP, which calculates the final trading price $\pi_{P2P}$ and the encrypted energy trade, $\e{V^{P2P}[u_k]}$, for each user $u_k$ in the P2P market, as in~\cite{erdayandi2022privacy}. 

During each SC, $\pi_{P2P}$ is publicly released. $\e{V^{P2P}[u_k]}$ is shared with related paired users for future calculations, and its hash is stored on the DLT for future verification. SMs measure their users' actual imported/exported electricity and transmit the encrypted version ($\e{V^{Real}[u_k]}$) to relevant users. The hash of this encrypted version is also stored on the DLT.

After sending and storing related data for billing, the calculation of bills among prosumers and consumers is performed in three stages in a privacy-preserving way. Firstly, individual deviations of users are calculated. Consumers calculate the individual deviations of prosumers and vice versa. Secondly, the total deviations of consumers and prosumers are calculated by six user selected from consumers and prosumers. Thirdly, statements (bills/revenues) of users are calculated.

To protect sensitive data such as energy consumed/traded, and individual energy deviations of households, our work utilises HE scheme to process data while preserving privacy. However, it is crucial to design the billing algorithm in such a way that it avoids any indirect leakage of private information despite the use of encryption. Traditional billing methods~\cite{thandi2022privacy,madhusudan2022billing} have the potential to expose confidential information by using individual deviations between actual and committed energy volumes to determine the ``conditions" in calculating bills. This enables inferences to be made about whether the actual electricity consumption volume is lower or higher than the committed data. To address this issue, we propose a privacy-preserving and accountable cost-splitting billing that uses total deviations of consumers and prosumers rather than individual deviations to determine billing conditions.

 In the event of a dispute, the referee requests the necessary data from households, as well as it retrieves the hash of the previously stored data from DLT (to ensure the accuracy of the data requested from households) to settle the dispute. In this case, the referee corrects erroneous computations of the pair of customer and prosumer whose calculations do not match each other and identifies the responsible party in the pair. The responsible party is penalised, incentivising them to act truthfully, which would otherwise result in penalties. Besides, the referee can directly calculate the supplier's balance since the calculations do not involve any confidential information.

 Finally, at the end of the month, final bills and revenues, and the balance of the supplier are released with the help of the referee and the private  homomorphic key of the supplier.

	\subsection{Technical Details of PA-Bill}
 \label{subsection:PA_technical}

	At the start of each billing period (e.g., a month), the following two steps (1-2) are carried out.  
        \subsubsection{Generation of Keys}
        The supplier generates a public-private HE (Paillier) key pair: $KGen_{pe}$($k$) $\xrightarrow{} PK_{sup}, SK_{sup}$. 

        \subsubsection{Matching customers and prosumers}
       The referee conducts a random matching process in which each consumer is paired with a list of prosumers and vice versa. 
       The number of users in the lists may exceed one or be zero in cases where $N_C > N_P$ or $N_C < N_P$, while  the lists contain only one user if $N_C = N_P$. Here, $N_C$ and $N_P$ denote the respective number of customers and prosumers. The function $M(u_k)$ returns the list of users that have been matched to the user $u_k$.

    At each SC, the following six steps (3--8) are carried out.
    
	\subsubsection{Transfer and Storage of P2P Traded Data} TP makes the P2P trading price public by storing it at DLT in plaintext. For each $u_k$, TP transmits homomorphically encrypted value of traded volume $\e{V^{P2P}[u_k]}$ to user $u_k$ and to users in $M(u_k)$. The privacy-preserving calculation of the encrypted traded values by user $u_k$ ($\e{V^{P2P}[u_k]}$) can be performed after the transmission of bids-offers in a homomorphically encrypted format. 
 It is assumed the TP has already calculated $\e{V^{P2P}[u_k]}$. Once the data has been transmitted to relevant parties, the TP also hashes the homomorphically encrypted traded volume of user $u_k$, i.e., $H{(\e{V^{P2P}[u_k]})}$, and stores the result at the DLT, together with a timestamp and ID of $u_k$.

	\subsubsection{Collection, Transfer and Storage of SM Data} 
 At the end of each SC, each SM measures the real volume of energy imported from (or exported to) the grid by their user, i.e., $V^{Real}[u_k]$, encrypts it with $PK_{sup}$ and hashes it, i.e., $H{(\e{V^{Real}[u_k]})}$. It then stores the hash value to DLT with timestamp and ID of $u_k$. The user SM also stores $\e{V^{Real}[u_k]}$ as well as sends it to the users in $M(u_k)$.

	\subsubsection{Calculation of Individual Deviations} \label{subsection:indev} 
in this step, each user $u_k$ calculates the individual deviations ($inDev$) from the volume of energy they committed for themselves and their corresponding matched users in $M(u_k)$ 
 (see Alg.~\ref{Algo:inDev}). To calculate $inDev$, each user $u_k$ subtracts their committed volume from the volume measured by their SM for themselves ($u_k$) and the users $m_l$ in $M(u_k)$. The calculations are carried out in homomorphically encrypted format. 
 The encrypted results $\e{inDev}$ and $\e{inDev_M}$ are sent to the referee.

After the referee receives the encrypted  individual deviations from users, it checks whether the computations have been done correctly. For each user and its matched user, the referee receives four encrypted results. The user $u_k$ provides its own encrypted result, ${\e{inDev[u_k]}}$, as well as that of its matched user. For the matched consumer $c_i$ and prosumer $p_j$, the referee checks if the calculated values are the same. In order to achieve this, the referee subtracts these two calculated values from each other in a homomorphically encrypted format. The result of this subtraction is then sent to the supplier who has the private key to perform homomorphic encryption operations. The supplier decrypts the result of subtraction and sends it back to referee. The referee checks whether the received value from the supplier is zero or not. If it is zero, it considers the calculations to be accurate and proceeds to store the hash of the resulting computation of user $u_k$ (not that of the matched user) in DLT along with the corresponding ID and timestamp of $u_k$, to facilitate future verification. Otherwise (if the received result is not zero), the referee intervenes to correct any erroneous calculations and identify the responsible party. To do so, the referee requests $\e{V^{Real}}$ and $\e{V^{P2P}}$ from the users, checks their correctness by hashing and comparing them with the previously stored hashes in blockchain by TP and SMs. If the encrypted data received from the users is accurate, the referee recalculates the inDev in encrypted format for $c_i$ and $p_j$, whose results were incorrect. Next, the referee follows the same process of subtracting the calculated values and having the result decrypted by the supplier to compare the recalculated outcome with the values obtained from $c_i$ and $p_j$. The referee then identifies the party that is accountable for the mismatch. 

	\begin{algorithm}[!t]
        
		\caption{Calculating Individual Deviations}
		\label{Algo:inDev}
		\SetAlgoLined
		\small
		\KwIn{$N_U$}
		\KwOut{$\e{inDev},\e{inDev_M}$}

		\For{$each~u_k$}{
			
			$\e{inDev[u_k]} \gets \e{V^{Real}[u_k]} - \e{V^{P2P}[u_k]};$\\

            \For{$each~m_l \pcin~M(u_k)$}{
                $\e{inDev_M[M(m_l)]} \gets \e{V^{Real}[M(m_l)]} - \e{V^{P2P}[M(m_l)]};$\\

            }
		}

	\end{algorithm}

  \subsubsection{Calculation of Total Deviations}
  
	To calculate total demand and supply deviations, the referee selects three consumers and three prosumers. Each consumer $c_i$ sends their respective $\e{inDev[c_i]}$ to the selected prosumers and vice versa. 
 Selected prosumers and consumers verify the received encrypted deviations by hashing and comparing them with stored hashes in DLT. Then, selected prosumers sum up $\e{inDev[c_i]}$ for each $c_i$ to calculate $\e{Dev_C^{Tot} }$ (eq. \ref{eq:devC}) and selected consumers do the same for each $p_i$, 
 (eq. \ref{eq:devP}).
 
 \begin{equation}
 \label{eq:devC}
\footnotesize \e{Dev_C^{Tot} } \gets \sum_{i=0}^{N_C-1} \e{inDev_C[c_i]}
\end{equation}

 \begin{equation}
 \label{eq:devP}
\footnotesize \e{Dev_P^{Tot} } \gets \sum_{j=0}^{N_C-1} \e{inDev_P[p_j]}
\end{equation}

After calculating $\e{Dev_C^{Tot}}$ and $\e{Dev_P^{Tot}}$, selected prosumers and consumers send them to a referee for verification. If the results match, the referee sends them to the supplier. 
The supplier then decrypts the results and makes them publicly available by storing ${Dev_C^{Tot}}$ and ${Dev_P^{Tot}}$ into DLT. If the results do not match, the referee corrects any erroneous calculations and identifies the responsible party. This is done by recalculating (eq.~\ref{eq:devC}) and (eq.~\ref{eq:devP}) in encrypted format after requesting and verifying the necessary data via DLT.

\begin{algorithm}[!t]
		
		\caption{Calculating Bills and Revenues}
		\label{algo:billing}
		\SetAlgoLined
		\footnotesize
		\KwIn{$ N_U, \e{V^{P2P}}, \e{V^{Real}}, Dev_C^{Tot}, Dev_P^{Tot}, \pi_{P2P}, \pi_{RT}$}
		\KwOut{$\e{Stat}, \e{Stat_M}$}		
		\For{$each~u_k$}{
    		\If{$ Dev_P^{Tot} = Dev_C^{Tot} $}{

    				$\e{Stat[u_k]} \gets \e{V^{P2P}[u_k]} \cdot \pi_{P2P} + \e{inDev[u_k]} \cdot \pi_{P2P}$\\
                    \For{$each~m_l \pcin~M(u_k)$}{
                        $\e{Stat[m_l]} \gets \e{V^{P2P}[m_l]} \cdot {\pi_{P2P}} + \e{inDev[m_l]} \cdot {\pi_{P2P}}$                
                    }
            }	

             \If{$Dev_P^{Tot} < Dev_C^{Tot}$}{

                $\e{Stat[u_k] }\gets \e{V^{P2P}[u_k]} \cdot \pi_{P2P} + \e{inDev[u_k]} \cdot \pi_{RT}$\\
                    \For{$each~m_l \pcin~M(u_k)$}{
                        $\e{Stat[m_l]} \gets \e{V^{P2P}[m_l]} \cdot {\pi_{P2P}} + \e{inDev[m_l]} \cdot {\pi_{RT}}$                
                    }

    		}

            \If{$Dev_P^{Tot} > Dev_C^{Tot}c$}{

                \eIf{ $u_k$ is a consumer}{
                    
    		         $\e{Stat[u_k]} \gets \e{V^{P2P}[u_k]} \cdot \pi_{P2P} + \e{inDev[u_k]} \cdot \pi_{P2P}$\\ 
        			 \For{$each~m_l \pcin~M(u_k)$}{
        				$\e{Stat[m_l]} \gets \e{V^{P2P}[m_l]} \cdot {\pi_{P2P}} + \e{inDev[m_l]} /Dev_P^{Tot}\cdot TotRev_P$\\
        			}
                }{
                    $\e{Stat[u_k]} \gets \e{V^{P2P}[u_k]} \cdot {\pi_{P2P}} + \e{inDev[u_k]} /Dev_P^{Tot}\cdot TotRev_P$\\
                    \For{$each~m_l \pcin~M(u_k)$}{
                        $\e{Stat[m_l]} \gets \e{V^{P2P}[m_l]} \cdot \pi_{P2P} + \e{inDev[m_l]} \cdot \pi_{P2P}$\\ 
                    }
                }

    		}

            $\e{stat^{Tot}[u_k]} \gets \e{stat^{Tot}[u_k]} + \e{stat[u_k]} $

             \For{$each~m_l \pcin~M(u_k)$}{
                $\e{stat_M^{Tot}[m_l]} \gets \e{stat^{Tot}[m_l]} + \e{stat[m_l]} $
             }

		}	
		
	\end{algorithm}

\subsubsection{Calculation of Bills and Rewards}

we present our proposed privacy-preserving and accountable universal cost-splitting billing model that employs total deviations instead of individual deviations to establish billing conditions. The proposed billing model is presented in Alg.~\ref{algo:billing}. The algorithm takes as input $\e{V^{P2P}}$, $\e{V^{Real}}$, $\pi_{P2P}$, $\pi_{RT}$ and $\pi_{FiT}$ and calculates the bills/revenues of consumers/prosumers. The algorithm outputs Statements $Stat[u_k]$, $Stat_M[u_k]$ for user $u_k$ and its matched users in $M(u_k)$, respectively. $Stat[u_k]$ indicates the bill of $u_k$ when $u_k$ is a consumer and it stands for the revenue of $u_k$ if $u_k$ is a prosumer. We have devised universal formulas such as $Stat[u_k]$ which is applicable to both consumers and prosumers.
 The algorithm works in three modes based on the difference between total deviations of consumers and prosumers, and proceeds as follows.

\textit{If $ Dev_P^{Tot} = Dev_C^{Tot} $}, prosumers have generated enough electricity to meet the demand of customers, resulting in a balanced P2P market. In this case, individuals can purchase the required energy from other households and sell their excess energy to other households at $\pi_{P2P}$ in addition to their commitments in the P2P market rather than relying on suppliers. Energy sharing between households to compensate for deviations is advantageous for both consumers and prosumers, as they can exchange energy at a price of $\pi_{P2P}$, which is higher than $\pi_{FiT}$ and lower than $\pi_{RT}$, compared to relying on suppliers to buy electricity at $\pi_{RT}$ and sell electricity at $\pi_{FiT}$. The statements for each user $u_k$ and for paired users in $M(u_k)$ are calculated between ln. 3-6 in the algorithm.

\textit{If $Dev_P^{Tot} < Dev_C^{Tot}$}, there is a shortage of electricity in the P2P market as prosumers have not generated enough electricity to meet customer demand. If there is a shortage of electricity that cannot be compensated by other users, the only option is to purchase it from the supplier at $\pi_{RT}$. Users with a shortage of electricity can buy it at this price, while households with a surplus can sell it at $\pi_{RT}$ instead of selling it to the supplier for $\pi_{FiT}$, which is advantageous for prosumers. In accordance with this, the statements for each user $u_k$ and for paired users in $M(u_k)$ are calculated between ln. 9-11 in the algorithm.

\textit{If $Dev_P^{Tot} > Dev_C^{Tot}$}, there is excess electricity in the P2P market as prosumers have generated more electricity than is needed to meet customer demand. In this case, consumers can purchase energy from prosumers at $\pi_{P2P}$ to compensate for their energy shortage due to deviation. The total revenue of the prosumers is distributed among them in proportion to the excess energy they provided. To calculate this, the total revenue generated by prosumers due to excess energy is first determined. Some of the excess energy is sold to consumers with a shortage of electricity at $\pi_{P2P}$, while the remainder is sold to the supplier at $\pi_{FiT}$. Therefore, the total revenue of prosumers, $TotRev_P$, can be calculated as  
\begin{equation}
    TotRev_P =(Dev_C^{Tot} \cdot \pi_{P2P} + (Dev_P^{Tot} - Dev_C^{Tot}) \cdot \pi_{FiT})    
\end{equation}

The total revenue $TotRev_P$ is distributed among the prosumers in proportion to ${inDev_P[u_k]} /Dev_P^{Tot}$. In accordance with this, Alg.~\ref{algo:billing} calculates statements for each user $u_k$ and for paired users in $M(u_k)$ between ln. 16-19, if $u_k$ is a consumer. Otherwise, the statements are calculated between ln. 21-24. 

At the end of the algorithm, statements are accumulated on $stat^{Tot}$ in encrypted format for $u_k$ and user in $M(u_k)$  assuming that $stat^{Tot}$ was set to zero before the first SC.

After each pair calculates their statements bilaterally, they send the results to the referee for verification. If the results do not match, the referee intervenes to correct any erroneous calculations and identify the responsible party. This is done by running Alg.~\ref{algo:billing} for the unmatched pairs after requesting and verifying the required data for computation via DLT.

\subsubsection{Calculating the of Balance of the Supplier}
The referee calculates the supplier's balance using only public information, and does so in a non-encrypted format.
In the case where $Dev_P^{Tot} = Dev_C^{Tot}$, $Bal_{sup}$ is set to zero ($Bal_{sup} \gets 0$) since there is no excess or shortage of electricity in the P2P market to compansate from the supplier.
If {($Dev_P^{Tot} > Dev_C^{Tot}$)}, there is excess energy in P2P market and the supplier purchases it at FiT price $\pi_{FiT}$, resulting in a negative balance for the supplier to pay. $Bal_{sup}$ is calculated as the negative product of the total excess energy $(Dev_P^{Tot} - Dev_C^{Tot})$ and $\pi_{FiT}$, i.e. 

\begin{equation}
    Bal_{sup} \gets -(Dev_P^{Tot} - Dev_C^{Tot})\cdot \pi_{FiT}
\end{equation}

If {($Dev_P^{Tot} < Dev_C^{Tot}$)}, there is a shortage of energy in P2P market that needs to be compensated by the supplier at retail price $\pi_{RT}$. $Bal_{sup}$ is calculated as the product of supplied energy $(Dev_P^{Tot} - Dev_C^{Tot})$ and $\pi_{RT}$, i.e. 

\begin{equation}
    Bal_{sup} \gets (Dev_C^{Tot} - Dev_P^{Tot})\cdot \pi_{RT}.
\end{equation}

At each SC,  the resulting $Bal_{sup}$ is accumulated to the total supplier balance 
except when the SC is equal to zero where $Bal^{Tot}_{sup}$ is set to $Bal_{sup}$.

	
	
	The next step is carried out at the end of each billing period.
	
	\subsubsection{Transfer and Announcement of Bills, Revenues and Supplier Balance}

The final accumulated monthly statements of households are not protected from the  supplier, as payments must be made, the referee sends encrypted statements consisting of bills and revenues to the supplier. The supplier then decrypts these statements using their HE private key and hashes and stores the decrypted version on the DLT system for future verification during the payment process. The supplier's balance is also hashed and stored on the DLT.

	\section{{Security, Privacy and Accountability Analysis} } 
	\label{section:security_privacy}

The PA-Bill protocol addresses the security concern of avoiding SPF by distributing the majority of calculations and data storage locally. 
It addresses privacy concerns by utilising HE to encrypt sensitive user data such as $V^{Real}$ and $V^{P2P}$, ensuring that sensitive information remains confidential during billing computations. In addition, the PA-Bill protocol employs a cost-splitting mechanism that utilises the total deviations of users rather than individual deviations to calculate billing modes. This method avoids indirect privacy leakage of individual deviations.
It employs Blockchain technology to create an unalterable record of the hashes of essential data necessary for billing computations. This ensures the verification and integrity of critical data, thereby enabling all parties to be held accountable for their actions during the billing process.

	\section{{Performance Evaluation} } 
	\label{section:performance_evaluation} 
In this section, we demonstrate that PA-Bill achieves computational efficiency without compromising privacy, accountability, or the ability to accommodate  user consumption variations. PA-Bill effectively addresses these critical aspects while maintaining a level of computational efficiency. We prove our claims through both theoretical analysis and experiments.

	\subsection{Theoretical Analysis}
	The time complexity of the method is mainly determined by the input parameters of Alg.~\ref{Algo:inDev} and Alg.~\ref{algo:billing}, which include the number of users ($N_U$). The time required to perform the algorithm grows depending on the input size. Specifically, the nested double loops in Alg.~\ref{Algo:inDev} and Alg.~\ref{algo:billing} lead to a quadratic time complexity of \bigO{n^2} for cases where in cases where $N_C > N_P$ or $N_C < N_P$, the time complexity is reduced to \bigO{n} with a single iteration in the inner loop when $N_C = N_P$ where each user has only one matched user. The time complexity of the calculations in eq.~\ref{eq:devC} and eq.~\ref{eq:devP} is \bigO{n}, where $n$ depends on the inputs $N_C$ and $N_P$, respectively.
	
	\subsection{Experimental Results}

   We evaluate the performance of PA-Bill by running simulations on a PC with Intel Core i5 CPU @ 2GHz CPU and 16GB of RAM to demonstrate its efficiency. We utilise the SHA3-256 algorithm for hashing and the Paillier cryptosystem for homomorphic encryption with 2048-bit keys. These operations were implemented using the Python libraries hashlib and phe, respectively. We utilised Ethereum network to prototype the blockchain platform.
    To deploy and test Ethereum for our project, we used Ganache\footnote{{https://www.trufflesuite.com/ganache}}, wrote smart contracts in Solidity\footnote{{https://solidity.readthedocs.io/en/v0.8.7/}}, and compiled them on Remix\footnote{{https://remix.ethereum.org/}}. To connect our project with the Ethereum network, we utilised the Python Web3\footnote{{https://web3py.readthedocs.io/en/stable/}} library. As we utilised existing tools to design the blockchain platform, we did not conduct a separate performance assessment of the platform itself. Our previous work~\cite{erdayandi2022privacy} is deployed as electricity trading platform, so we do not reevaluate it in this context either. Instead, our primary focus lies in evaluating the performance of the privacy and accountable billing model.

The billing model simulations were conducted on a sample of 500 users, consisting of 250 consumers and 250 prosumers. We measured PA-Bill's execution time (ET) for computationally intensive components in two scenarios: worst-case (every household makes an incorrect bill calculation (unintentionally or maliciously), thus requiring an intervention from the referee) and best-case (all households make correct calculations, hence no referee intervention is deployed).
The SC is set to be one hour. Table~\ref{table:runtime} demonstrates the average execution time per SC for PA-Bill components, computed over a one-month billing period comprising 720 SCs (24 SCs per day). The execution time which results in milliseconds for both worst-case and best-case scenarios, tested with a large group of $500$ users, indicate that our proposed billing protocol offers a computationally efficient solution for PA-Bill.

	\begin{table}[]

 \caption{Execution time results per settlement cycle. }
  \label{table:runtime}
 \centering
\begin{tabular}{l c c }
\toprule
\textbf{Calculation step }                    & \textbf{Best-case ET}   & \textbf{Worst-case ET} \\
\midrule

Individual Deviations & 23.84 ms     & 48.64 ms                  \\
 Total Deviations      & 69.25 ms   & 246.23 ms                     \\
 Bills and Rewards     & 25.76 ms & 50.15 ms     \\  
\bottomrule
\end{tabular}
\end{table}

	\section{{Conclusion} } 
	\label{conlcusion_future_work}

In this work, we proposed PA-Bill, a privacy-preserving and accountable billing protocol that addresses security, privacy, and accountability issues in P2P markets at the billing and settlements stage. PA-Bill utilises a universal cost-splitting billing model, local semi-decentralised calculation, and Homomorphic Encryption for privacy protection. Blockchain technology is deployed for accountability mechanisms that resolve conflicts during billing calculation. PA-Bill is evaluated on a community of 500 households. In our future work, we plan to investigate network constraints.

	\bibliographystyle{IEEEtran}
	\bibliography{sample.bib}

\end{document}